\documentclass[letterpaper, 10 pt, conference]{ieeeconf}  

\IEEEoverridecommandlockouts                              

\usepackage{graphicx} 
\graphicspath{{images/}}  
\usepackage{epsfig} 
\usepackage{url}
\usepackage{booktabs}
\usepackage{hyperref}
\usepackage{color}
\usepackage{stfloats}
\usepackage{algpseudocode}
\usepackage{algorithm}

\title{\LARGE \bf
A Miniaturised Camera-based Multi-Modal Tactile Sensor
}

\author{Kaspar Althoefer$^{1 * }$, Yonggen Ling$^{2 *}$, Wanlin Li$^{3 *}$, Xinyuan Qian$^{4}$, Wang Wei Lee$^{2}$, Peng Qi$^{5}$
\thanks{$*$ Authors with equal contribution.
    {\tt\small }}%
\thanks{$^{1}$ The Centre for Advanced Robotics @ Queen Mary (ARQ), Queen Mary University of London, London, United Kingdom
	{\tt\small }}%
\thanks{$^{2}$ Tencent Robotics X Lab, Shenzhen, China {\tt\small }}%
\thanks{$^{3}$ Beijing Institute for General Artificial Intelligence (BIGAI), Beijing, China {\tt\small }}%
\thanks{$^{4}$ The Department of Computer and Communication Engineering, University of Science and Technology Beijing, Beijing, China {\tt\small }}%
\thanks{$^{5}$ Department of Control Science and Engineering, College of Electronics and Information Engineering, Tongji University, Shanghai, China {\tt\small }}%
\thanks{This work was done during Wanlin Li's internship at Tencent.
{\tt\small }}%
}

\begin{document}

\maketitle
\thispagestyle{empty} 



\begin{abstract}
In conjunction with huge recent progress in camera and computer vision technology, camera-based sensors have increasingly shown considerable promise in relation to tactile sensing. In comparison to competing technologies (be they resistive, capacitive or magnetic based), they offer super-high-resolution, while suffering from fewer wiring problems. The human tactile system is composed of various types of mechanoreceptors, each able to perceive and process distinct information such as force, pressure, texture, etc. Camera-based tactile sensors such as GelSight mainly focus on high-resolution geometric sensing on a flat surface, and their force measurement capabilities are limited by the hysteresis and non-linearity of the silicone material. In this paper, we present a miniaturised dome-shaped camera-based tactile sensor that allows accurate force and tactile sensing in a single coherent system. The key novelty of the sensor design is as follows. First, we demonstrate how to build a smooth silicone hemispheric sensing medium with uniform markers on its curved surface. Second, we enhance the illumination of the rounded silicone with diffused LEDs. Third, we construct a force-sensitive mechanical structure in a compact form factor with usage of springs to accurately perceive forces. Our multi-modal sensor is able to acquire tactile information from multi-axis forces, local force distribution, and contact geometry, all in real-time. We apply an end-to-end deep learning method to process all the information.

\end{abstract}

\section{INTRODUCTION}

For humans and robots alike, the sense of touch is fundamental to the ability to understand, interpret and interact with the environment. As a consequence, the development of accurate force and tactile sensing is a key goal in robotics - indeed successful robot-environment interaction is reliant on end-effector sensors providing the robot with the relevant feedback. Despite this, the use of tactile sensing remains somewhat limited, and its development in recent years has been relatively slow in comparison with the dramatic developments in computer vision. One of the principal reasons for this is that from a hardware perspective, current robotic tactile systems fall short of their human equivalents in terms of their efficacy and, equally importantly, their compactness.


\begin{figure}[h]
	\centering
	\includegraphics[width=0.88\columnwidth]{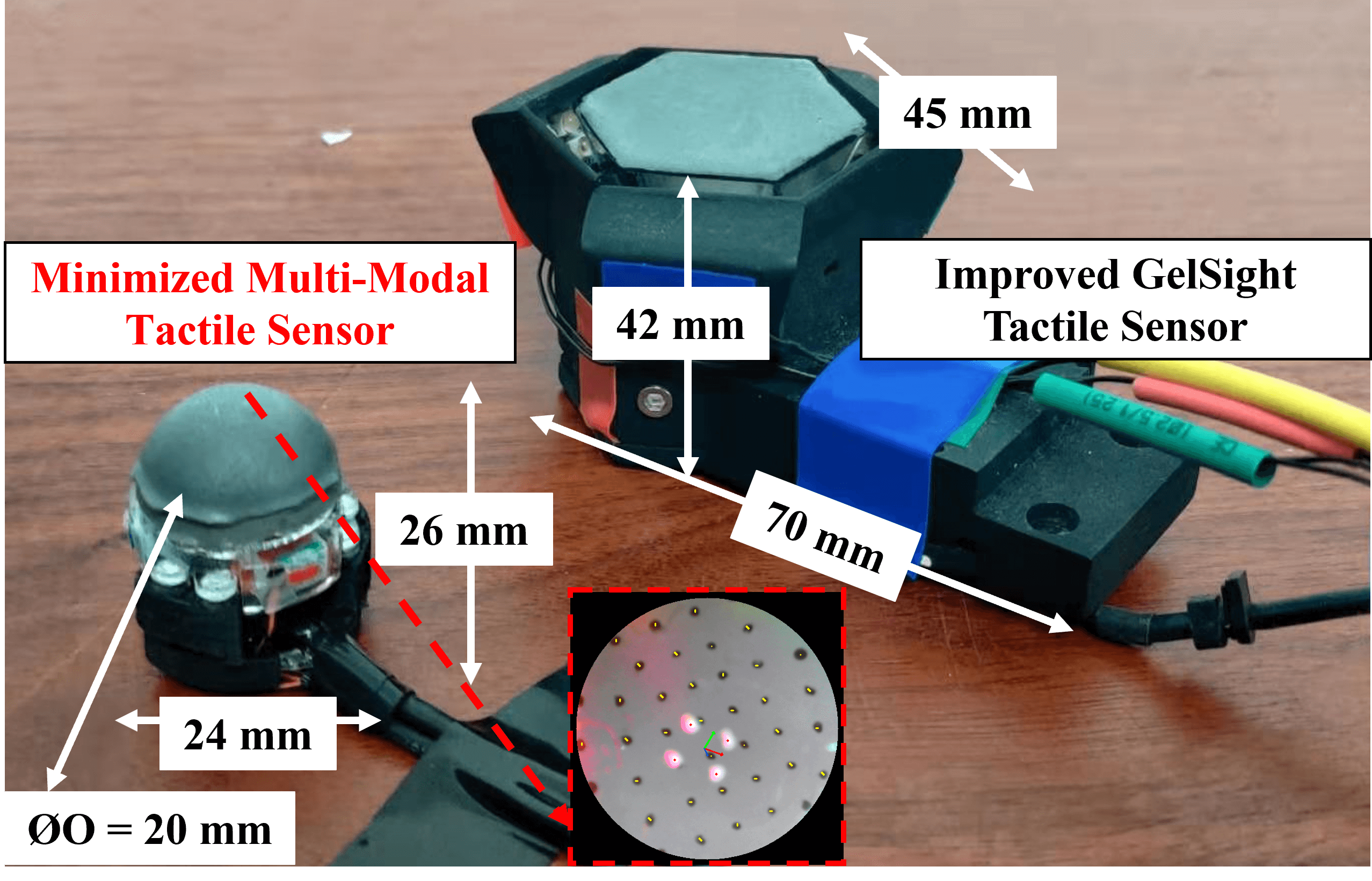}
	\caption{Our miniaturized multi-modal sensor next to an improved GelSight tactile sensor \cite{dong2017improved}. Compared to the GelSight, our sensor uses (1) a dome-shaped elastomer surface with uniform markers; (2) an improved illumination system with small diffused LEDs; (3) a force-sensitive mechanical structure in a compact form factor.}
    
	\label{fig:Figure1}
\end{figure}

The human tactile system is highly complex,  and capable of integrating a wide range of sensing characteristics. The glabrous skin is endowed with a high density of epidermal mechanoreceptors (Pacinian corpuscles, Meissner’s corpuscles, Ruffini corpuscles, and Merkel cells), arranged at different depths within the epidermis \cite{dahiya2009tactile}. Two of these, in particular, are responsive to force and texture perception:

\begin{enumerate}
    
    \item  Ruffini corpuscle:
    responsive to low-resolution multi-directional static forces based on skin stretched near joints (intrinsic tactile sensing);

    \item  Merkel cells:
    responsive to high-resolution skin deformation, sustained pressure and texture perception (extrinsic tactile sensing).
    
\end{enumerate}


Inspired by the human tactile system, sensors using many different technologies have been proposed, in which a variety of functional sensing elements are integrated into the robot end-effector to provide the requisite feedback. Among them, camera-based tactile sensors \cite{cui2021self, trueeb2020towards, du2021high, gomes2020geltip, abad2021haptitemp, do2022densetact, nozu2018robotic} show certain advantages, among them super-high-resolution as each pixel from an image can be regarded as a "tactel" (tactile element), and ease of use as they use fewer cables than other capacitive-based, resistive-based, OFETs and OECTs tactile array sensors.

\begin{figure*}[t]
	\centering
	\includegraphics[width=1.79\columnwidth]{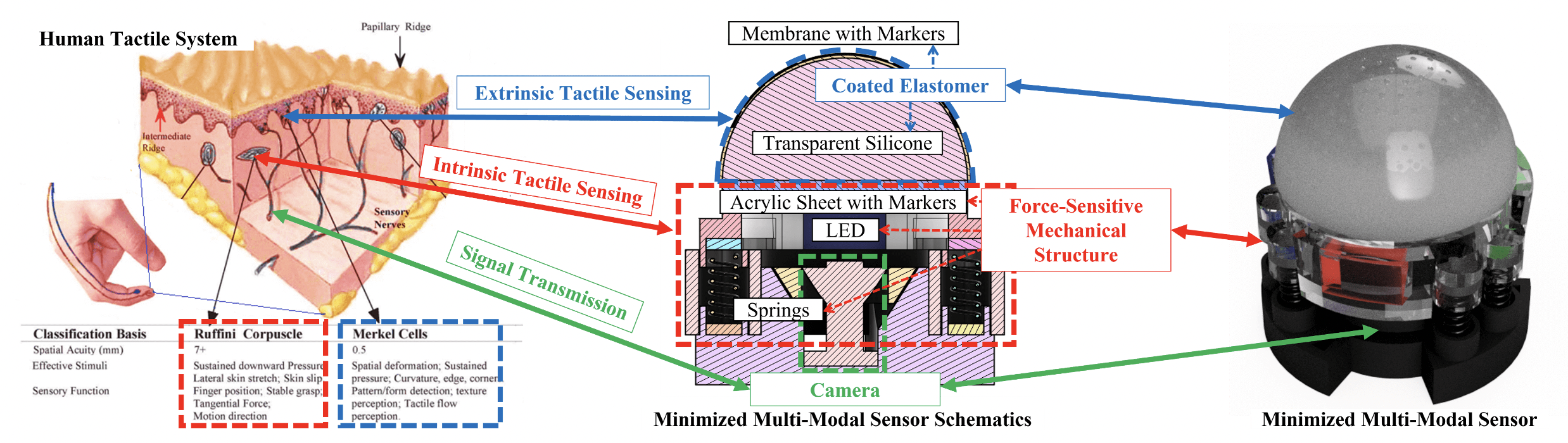}
	\caption{Comparison of our miniaturised sensor (middle and right side) to a human tactile system (left side). With an arrangement of a domed-shaped coated elastomer (for tactile array sensing and texture perception, similar to the functionality of the Merkel cells in the human tactile system) bonding upon a force-sensitive mechanical structure (for force sensing, similar to the functionality of the Ruffini corpuscle), a camera (similar to the functionality of the sensory nerves) at the bottom can directly capture the deformation of the two components, therefore transferring both extrinsic and intrinsic tactile information to the computer for further processing.}
	\label{fig:Figure2}
\end{figure*}

In this paper, we present an innovative design, see Figure~\ref{fig:Figure1}, that offers better functionality and greater compactness than earlier camera-based tactile sensors. Initially, we outline the design and fabrication processes for the compact multi-modal sensor (see Section III). We then evaluate its properties (see Section IV) and demonstrate its potential uses by applying an end-to-end learning method to a binary hardness classification task.

\begin{table*}[hbp]
\centering
\caption{Summary of the GelSight and GelSlim families and our sensor}
\label{tab:Chart1}
\begin{tabular}{|c|c|c|c|c|c|c|c|}
\hline
\textbf{Tactile Sensors} & \textbf{Year} & \textbf{End-effector} & \textbf{Illumination} & \textbf{Force} & \textbf{Shape} & \textbf{Remarks} \\ \hline

TacTip \cite{ward2018tactip}                             & 2009          & To robot arm           & White LEDs      & Force distribution            & Round            & Use pins to measure force distribution    \\ \hline

GelSight \cite{yuan2015measurement}                             & 2015          & Robot gripper           & RGBW LEDs      & Force distribution            & Flat            & Use photometric stereo for reconstruction     \\ \hline

FingerVision \cite{yamaguchi2017implementing}                    & 2017          & Robot gripper      & No LEDs                & Force distribution            & Flat      & Use environment light for observation  \\ \hline

GelSlim \cite{ma2019dense}                          & 2019          & Robot gripper  & RG LEDs               & Force distribution            & Flat         & Use IFEM to measure force distribution\\ \hline

GelTip \cite{gomes2020geltip}                   & 2020          & Robot gripper     & RGB LEDs              & No             & Round          & Omnidirectional observation\\ \hline

DIGIT \cite{lambeta2020digit}                   & 2020          & Robot hand     & RGB LEDs               & Force distribution             & Curved       & Compatible to robot fingertip \\ \hline

GelStereo \cite{cui2021self}                   & 2021          & Robotic gripper     & White LEDs               & Force distribution             & Flat       & Use stereo camera for reconstruction \\ \hline

Our Sensor                   & 2022          & Robot hand     & Diffused RGB LEDs               & Multi-axis force             & Round       & Use force-sensitive structure \\ \hline

\end{tabular}
\end{table*}

\section{Related Work}
Research into camera-based tactile sensors has become commonplace alongside developments in small compact cameras and computer vision technology. A camera is usually placed inside the sensor to observe the deformation of the sensing medium within different frames. Many sensors have been produced in recent years focusing on different aspects, among them multi-axis force measurement, pressure sensing, and geometric perception. A typical example is the GelSight and GelSlim family, designed to estimate geometry and force information. In 2009, Adelson et al. introduced an elastomeric sensor \cite{johnson2009retrographic} that can reconstruct the contact surface's shape and texture. This was a precursor to GelSight, which was able to obtain ultra-high-resolution tactile images beyond even human tactile perception. Several vision-based tactile sensors were then produced and a comparative summary of some of their characteristics is shown in Table \ref{tab:Chart1}. It can be seen that the evolution of these sensors includes a shift from bulkier formats down to dimensions compatible with small grippers and robot hands. Different illumination methods have been explored and employed for better image quality; markers have been uniformly added to the top of the silicone surface to measure force information (although as yet never onto a spherical surface). Different shape profiles have also been produced (flat surface for grippers and rounded shape for robotic hands).


In this work, we present a new miniaturised camera-based multi-modal tactile sensor. We mainly focus on the miniaturisation of form factors, compactness, sensor manufacture, and illumination quality. Our sensor is capable of both extrinsic and intrinsic tactile sensing due to the use of a rounded silicone elastomer (with markers) and a force-sensitive mechanical structure within the sensor body. It is worth noting that using a relatively rigid structure provides for better multi-axis force measurement than having markers on soft materials due to its hysteresis. Our sensor is in a compact form factor with a dome surface that can be mounted onto robotic grippers and hands. Because the underlying sensing element of our new sensor is a camera, data that is rich in information can be acquired and exploited by computer vision-based learning algorithms. We enhance the internal illumination system by means of  diffused LEDs that provide a uniform lighting environment for better image quality.

\section{Sensor Design and Fabrication}

This section provides further detail on the design and fabrication of the miniaturised compact multi-modal tactile sensor, with a dome-shaped Lambertian surface (with uniform markers), a force-sensitive spring-like mechanism structure, and a tiny endoscope camera (see Figure~\ref{fig:Figure2}).

\subsection{Tactile Sensor Design Principle}

An ideal artificial tactile system needs to be compact and versatile. The improved GelSight sensor \cite{dong2017improved} is designed to provide comprehensive information on force and geometric characteristics when in contact with the external environment. However, it does have limitations. Firstly, the device is relatively bulky, and therefore incompatible with robot hands and small grippers used in manipulation tasks. Secondly, its sensing surface is relatively flat, which restricts its use in certain situations. For example, \cite{romero2020soft} demonstrates cases where the size of the contact patch is reduced to a point contact during exploration, thereby leading to an inconsistency in measurement results. Thirdly, although GelSight is intended to produce homogeneous illumination,  SMD LEDs are usually spotty focus light sources. Therefore, they require some physical distance for the lighting path in order to achieve wave uniformity. Additionally, GelSight adds multiple markers onto an elastomer surface for force estimation, which can generate hysteresis leading to lower accuracy.

To overcome these challenges, we propose the following design adaptations for the new sensor design (see Figure~\ref{fig:Figure3}). The innovations of the proposed sensor are as follows. 

\begin{enumerate}
    
    \item  \textbf{Miniaturisation:}
     Small tactile sensors have the potential to handle small objects. Our sensor ($24\;mm * 24\;mm * 26\;mm$) is designed to fit onto small end-effectors. The sensor is composed of (1) a small coated elastomer with a diameter of $20\;mm$; (2) a 3D printed miniaturized structure, with small compression springs and magnets; (3) a $0.8\;mm$ thick PCB with $1.2\;mm$ height surface-mounted LEDs; (4) a $7.5\;mm$ diameter ultra mini CMOS camera with a $160^{\circ}$ lens.

    \item  \textbf{Thumb-like Structure:}
    The human finger has a curved surface to facilitate daily tasks. We designed the sensor elastomer to be dome-shaped (like a thumb) rather than a flat surface. We also added uniformly distributed black markers to the curved surface to estimate the local force distribution. The coated elastomer is bonded to the spring-like mechanism structure (for multi-axis force measurement) to provide a compact form factor, comparable to a human thumb, for multi-modal signal acquisition.
    
    \item  \textbf{Uniform Illumination:}
    The illumination system in our fingertip sensor has been redesigned. We applied rough-surface materials around the sources to diffuse LEDs into smooth glow light for uniform lighting on the curved Lambertian reflective surface. In doing so we shortened the distance of the light path to ensure a compact minimised sensor structure.
    
    \item  \textbf{Durability and Reliability}
    In order to increase the durability and reliability of the elastomer, we followed the methods in \cite{romero2020soft}  to manufacture the coating layer. 
    
\end{enumerate}

\subsection{Tactile Sensor Fabrication}

\begin{figure}[t]
	\centering
	\includegraphics[width=1\columnwidth]{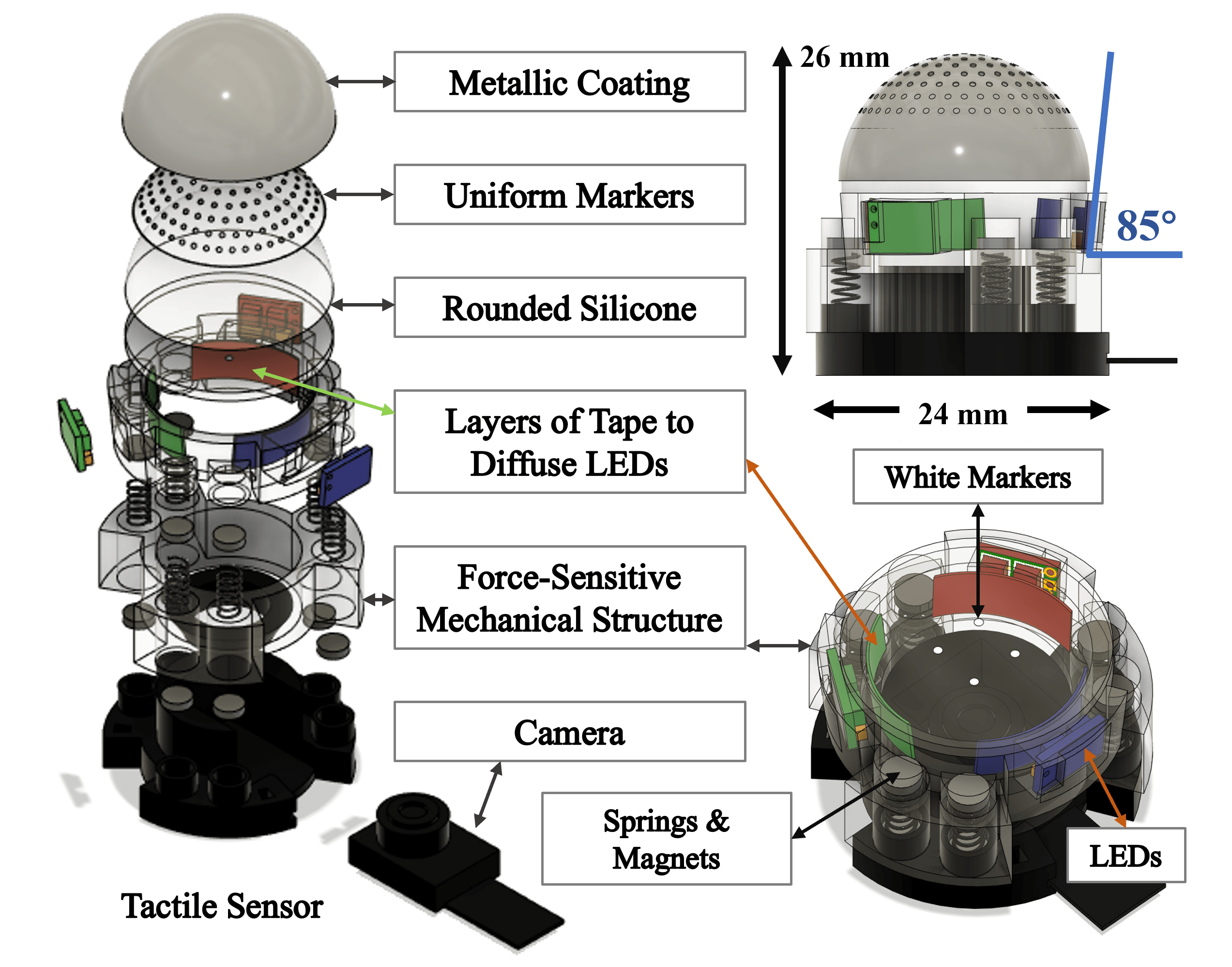}
	\caption{Exploded view of the miniaturised tactile sensor demonstrating the internal components of the coated elastomer, force-sensitive mechanical structure, the illumination system, and the camera.}
	\label{fig:Figure3}
\end{figure}

\begin{figure}[b]
	\centering
	\includegraphics[width=0.7\columnwidth]{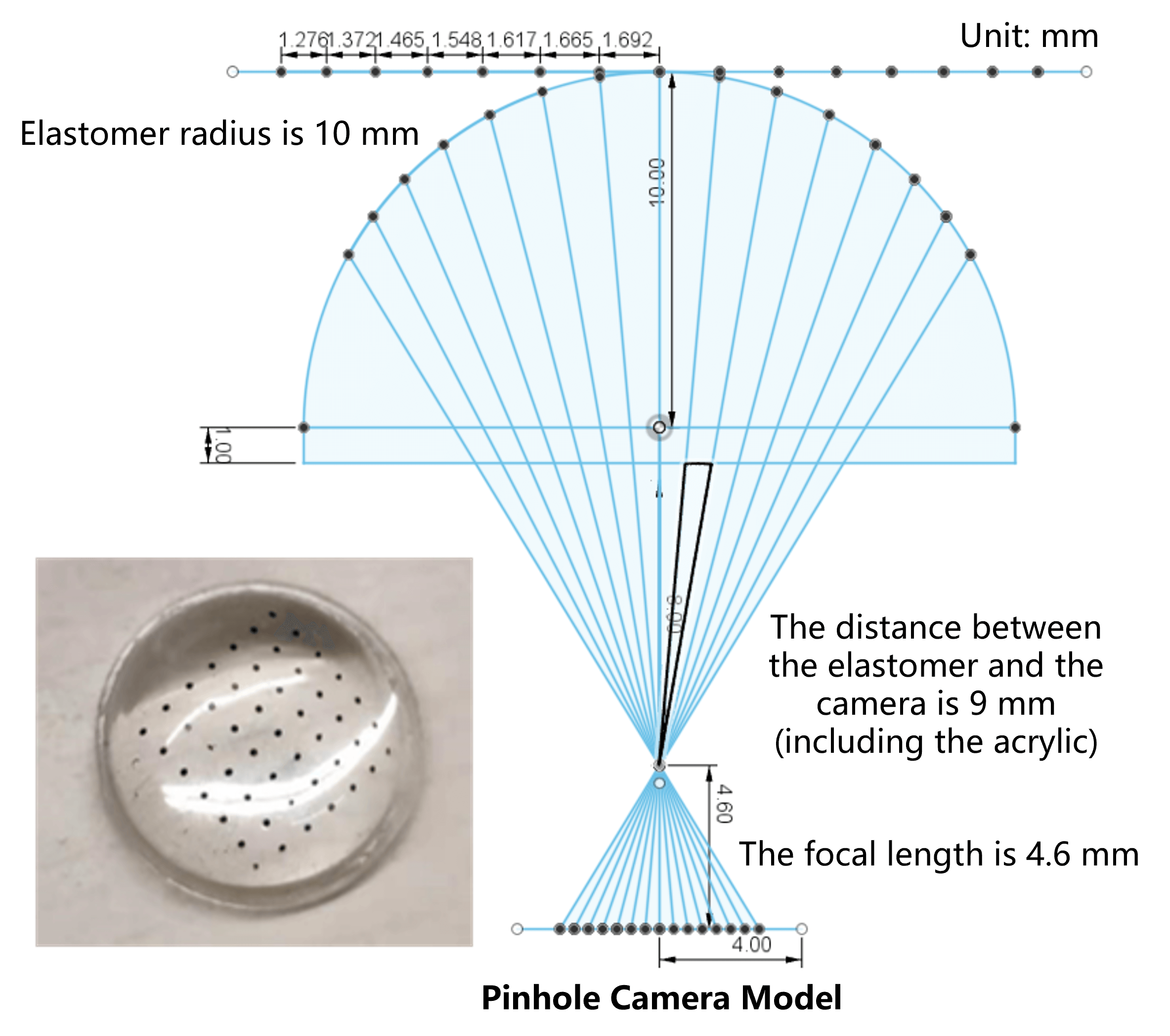}
	\caption{A pinhole camera model for generating uniform dots pattern displayed in the camera frame. We calculate the distance between each dot on the curved surface given the physical size of the sensor and camera parameters. We use pad printing to transfer the dots pattern to the rounded transparent silicone surface.}
	\label{fig:Figure4}
\end{figure}


We followed \cite{dong2017improved} to manufacture the elastomer with a Lambertian reflective membrane and black markers. However, Adding tiny markers onto a rounded elastomer surface is hard due to the fact that oil-based silicone ink takes a long time to dry and the printed markers can easily be erased from the surface. Therefore, we used pad printing to transfer the black dotted pattern from the soft silicone pad to our transparent silicone. Moreover, in order to observe evenly spaced markers in the pixel frame, we built a pinhole camera model to calculate the distances between pattern dots so that the dots on the rounded surfaces would be ideally displayed in the camera capture (see Figure~\ref{fig:Figure4}).


\begin{figure*}[t]
	\centering
	\includegraphics[width=2\columnwidth]{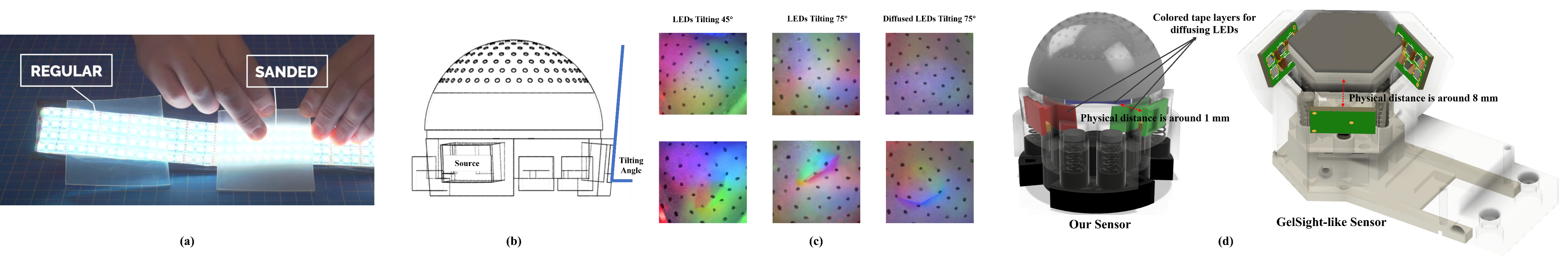}
	\caption{(a) Comparison of the influence of light diffusion material (with different surfaces) on the light source \cite{diffusedLED} . (b) Schematic design of the sensor illumination system. (c) Comparison of the influence of the light source tilting angle and the diffusion material. We aim to obtain equal lighting conditions on the sensing surface. (d) Our illumination system shortens the lighting path (compared to the GelSight-like sensor), resulting in a compact form factor.}
	\label{fig:Figure5}
\end{figure*}

We redesigned the illumination system by utilising diffused LEDs to shorten the light path distance. GelSight aims to obtain equal lighting conditions for each pixel across the sensing surface but the RGB LEDs are allocated in separate corners. It therefore requires fine-tuning of both the distance and the irradiation angle of the LED arrays (Figure \ref{fig:Figure5}(d)). Since rough-surface materials can change the angle of the light beam, resulting in light diffusion \cite{diffusedLED} (Figure \ref{fig:Figure5}(a)), we stuck tapes of different colours on the lighting path of each source, and adjusted the inclination angle of the source (as shown in Figure~\ref{fig:Figure5} (d)) to produce homogeneous lighting conditions on the dome surface. The influence of the inclination angle and the diffuse material can be seen in Figure~\ref{fig:Figure5} (b) and (c). In this work, we chose LUXEON CZ RGB LEDs on account of their compact size and high luminous flux. All LEDs were tilted by $85^{\circ}$ with respect to the elastomer bottom plane. RGB coloured tapes were sealed after installing the PCBs.

\begin{figure}[htp]
	\centering
	\includegraphics[width=0.98\columnwidth]{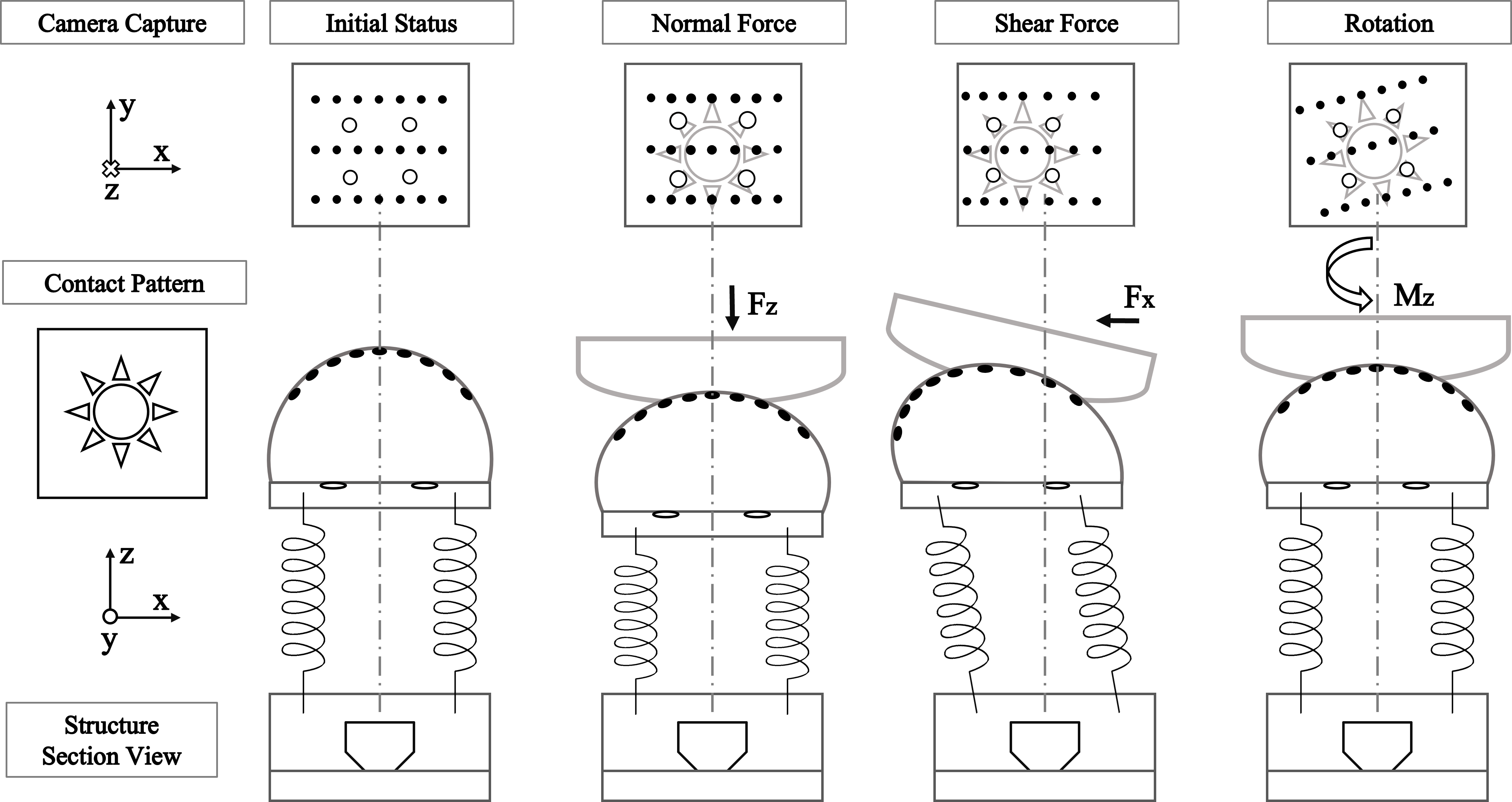}
	\caption{Working principle of the sensor. The top row depicts the camera view under different contact conditions. The movement of the white markers is derived from the spring compression. By calculating the 6-axis spatial pose of the plane composed of the white markers, we can derive 6-axis force/torque applied to the sensor. Black dots are used to illustrate the local force distribution. Contact patterns can be directly observed from the capture.}
	\label{fig:Figure6}
\end{figure}

Our sensor uses a spring-like mechanism structure to measure multi-axis force and torque. Four tiny white markers are painted on the top surface of the acrylic sheet (see Figure~\ref{fig:Figure3}) to track the overall deflection of the structure (see Figure~\ref{fig:Figure6}). The upper and lower platforms are 3D printed from transparent VeroClear material. The compression springs are $5\;mm$ in length and $3\;mm$ in diameter, with a $0.3\;mm$ wire diameter. As the springs cannot easily be glued to the platform (the wire is too thin), twelve $3\;mm$-diameter magnets (glued to platforms) are used to connect the springs (via magnetic attraction). An ultra mini CMOS colour UVC camera (MISUMI Group Inc.) is used and fixed to the lower platform. 

\section{Sensor Evaluation}

Using this tactile sensor, we can obtain three modalities from a single image in real time without too much computation (30 Hz in Python-OpenCV). They are (1) local force distribution from the motion of the black markers; (2) multi-axis force/torque information from the spatial translation and rotation of the four white markers; (3) geometric information from the tactile image. The detailed methodology for obtaining data for each modality is outlined below.



The use of a wide-angle lens in our sensor ensures a full view of the entire elastomer surface. It does however also cause considerable image distortion. Therefore, when initialising the sensor, the first step is to apply a correction function based on the camera’s intrinsic matrix $K$ and the distortion coefficients $(k_1,\;k_2,\;p_1,\;p_2,\;k_3)$, both of which are acquired from the standard pinhole camera calibration. The initial frame is saved and set as the reference frame $F_0$. For each subsequent frame, we then apply Gaussian blur to reduce the image noise. In addition, we apply a circular mask with a diameter equal to the height of the frame to remove unwanted areas (the obtained frame is $F$) and, then, apply a sharpening function to enhance the edges of both black and white markers. The RGB frame is transformed into a greyscale image $F_{gray}$, as follows:

\begin{figure*}[htp]
	\centering
	\includegraphics[width=1.9\columnwidth]{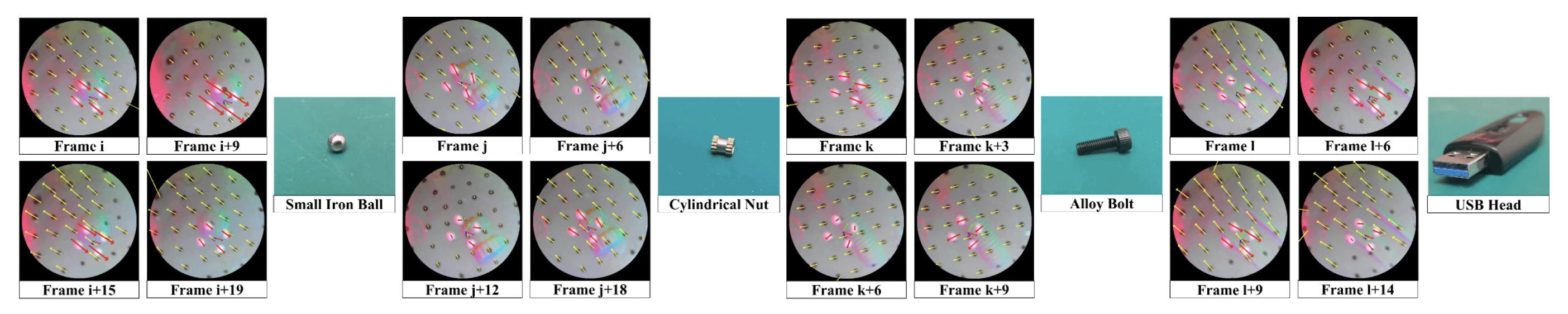}
	\caption{Examples of pressing different objects on top of our sensor's surface at different frames. From left to right are: A small iron ball; a cylindrical nut; an alloy bolt; a USB head. In the sensor capture, the tactile imprint shows the geometric information of the contact objects; the yellow arrows illustrate the movement of black markers between adjacent frames, indicating the force distribution; the red arrows illustrate the movement of white markers (the spatial pose is represented by the RGB coordinate) between the current and the reference frame, indicating the net force and torque.}
	\label{fig:Figure7}
\end{figure*}

\begin{itemize}
\item For force distribution detection, we extract black markers from the previous frame $F_{gray0}$ with a low threshold, and apply morphological transformations to remove noise. We then find the connected components of each marker in the image and save its centroid coordinate $p_0$. For the subsequent frame, we apply an optical flow method with $F_{gray0}$, current greyscale frame $F_{gray1}$, and $p_0$ to calculate the new feature points $p_1$. This allows us to track the movement of the black markers between adjacent frames; this movement is then shown by yellow arrows overlaid onto the image (see Figure~\ref{fig:Figure7}).       

\item For multi-axis force and torque evaluation, we extract white markers from the frame $F_{gray}$ with a high threshold. We then apply the same morphological transformations and connected components function (as done above) to locate each centroid coordinate $P_1$. We apply the SolvePnP method with the markers' coordinates in 3D space $S_1$ (measured in advance), $P_1$, $K$, and zero distortion coefficients to obtain the 6D pose estimation $P$ (three translation values and three Euler angles). Through the design of the force-sensitive spring-like structure, the multi-axis force/torque value is linearly correlated to $P$. We plot the force changes using red arrows (using $F_0$ as the reference coordinate). The measured maximum load force is $17\;N$. An in-depth evaluation of this approach can be referred from \cite{li2020f1}.

\item For geometric information, we use frame $F$ which contains depth information from the acquired RGB colours in the captured frame \cite{johnson2009retrographic}.

\end{itemize}
 

\begin{table}[b]
\centering
\caption{Evaluation of the trained model ResNet18-GRU}
\label{tab:Chart2}
		\begin{tabular}{cccccc}
			\toprule
			 \textbf{Training} & \textbf{Testing} & \textbf{Epochs} & \textbf{CE Loss}  & \textbf{-}  & \textbf{Precision (\%)}   \\ 
			\midrule
			 24,000 & 6,000 & 50 & 0.031  & -  & 99.054  \\ 
			\bottomrule			
		\end{tabular}
	\end{table}


\begin{figure}[!t]
	\centering
	\includegraphics[width=1\columnwidth]{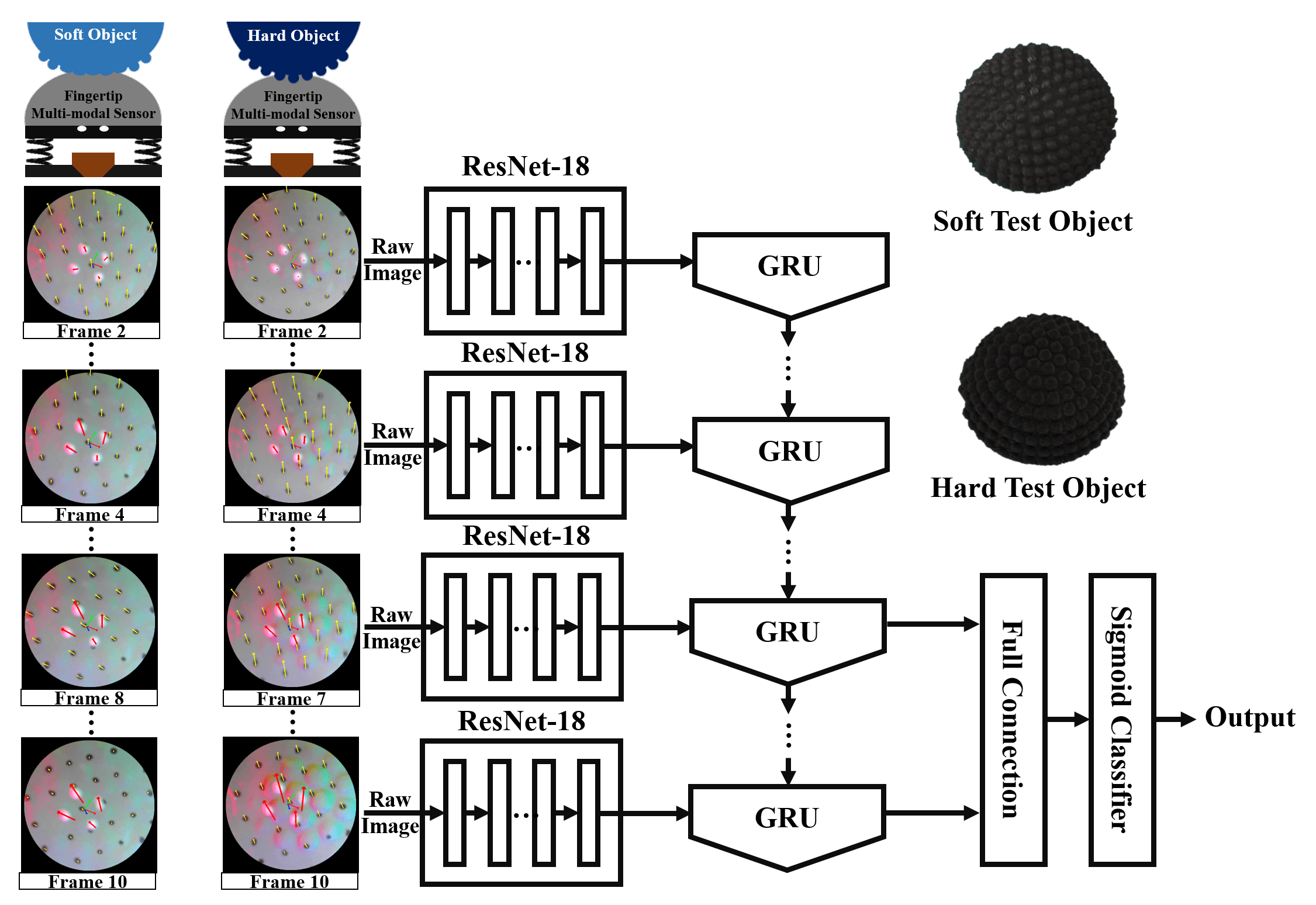}
	\caption{An experiment where our tactile sensor contacts hard and soft objects of the same appearance. We use a network architecture consisting of the ResNet-18 and a bi-directional GRU (gated recurrent units) to classify the hardness (soft or hard) of the object. We evenly choose 10 frames as the input and use the output of the last five frames to predict the class.}
	\label{fig:Figure8}
\end{figure}

\section{Test and Experiment}

With the above manual feature extractions, we tested and pressed objects of different shapes on top of our tactile sensor, as shown in Figure~\ref{fig:Figure7}. The sensor can perceive local force distribution (in yellow arrows), force estimation (in red arrows), and contact geometry in real-time. Deep learning shows advantages in end-to-end processing and the GelSight tactile sensor can measure the object's hardness based on the sensor's raw images \cite{yuan2017shape}. However, the GelSight sensor offers reduced accuracy when measuring net forces due to the hysteresis of the soft material. Our sensor obtains rich information (multi-axis force, force distribution, and geometry signals) from images of both the deformed elastomer and the force-sensitive structure, which makes it suitable for an end-to-end approach. 

One advantage of a camera-based tactile sensor is that it can perceive aspects of an object's characteristics that cannot be ascertained by vision alone. We set up an experiment to distinguish between two objects with the same appearance but with different levels of hardness. A typical hardness test is performed by pressing an indenter (with a given load) into a test object and measuring the depth of indenter penetration. The object's hardness is then determined by the relationship between the applied force and the object's deformation. In this experiment however, we used a neural network to distinguish between the two objects (see Figure~\ref{fig:Figure8}) based on a sequence of raw images from our sensor. The two test objects are of different hardness: the soft one is cast with Ecoflex 00-50, the hard one is 3D printed (30 Shore A).


\subsection{Neural Network Design}

The design of the neural network architecture is presented in Figure~\ref{fig:Figure8}, and the code will be made publicly available. The network input is a sequence of raw sensor images (without arrows and coordinate symbols) representing the contact process. The input images (containing both force and tactile information) are propagated through the ResNet-18 convolutional network, from which the features are extracted in the fully-connected layer ($fc$) with a dimension of 128. The encoded image features are then propagated through the bi-directional gated recurrent units (GRU) neural network to model the sequence information. 
Here we select the latent features of the last five frames as the inputs to the Sigmoid classifier for predicting the hardness (two classes with $0$ standing for soft and $1$ standing for hard, respectively).



\subsection{Experimental Results}
To evaluate our sensor's ability to distinguish between two objects, we pressed objects onto the top of our sensor with random force, and recorded the sensor captures. We collected $500$ videos, each one containing $60$ frames. By selecting $10$ frames from each video we form a training data-set containing about $24,000$ frames and $6,000$ frames for validation. We trained the model of both ResNet-18 and GRU using the stochastic gradient descent (SGD) as the optimiser, with a learning rate of $0.001$. We used cross-entropy loss, and we trained the model for 50 epochs within a few hours and selected the model with the best performance for final validation. We validated our model on test frames of both objects, and the result is shown in Table \ref{tab:Chart2}. Our sensor achieved an accuracy of $99.054\%$, which proves it can well distinguish between soft and hard objects of similar appearance, using both force and geometric information.


\section{Conclusion and Future Work}
In this paper, we introduced a new miniaturised camera-based multi-modal tactile sensor, that can perceive multi-axis force, local force distribution, and high-resolution geometry. The sensor is dome-shaped, with a marked interaction surface, a spring-like mechanism, and, in comparison to earlier tactile sensors, an improved illumination system. We also demonstrated how a neural network can be used alongside the sensor to distinguish between objects of the same appearance but of differing hardness.

In the future, we plan to expand our work on object hardness estimation using an extensive data-set of objects, and demonstrate better end-effector manipulation capabilities, once the sensor is installed.

\section*{Acknowledgement}

We thank Yu Zheng, Dongsheng Zhang, Zunran Wang for in-depth discussions, and Mish Toszeghi for proofreading. Qi is supported by the National Natural Science Foundation of China under Grant 62273257; and Shanghai Science and Technology Development Funds (20QC1400900). Althoefer is supported by the Alan Turing Institute, UK and the National Centre for Nuclear Robotics, UK (EP/R02572X/1). For the purpose of open access, the author(s) has applied a Creative Commons Attribution (CC BY) license to any Accepted Manuscript version arising.



\bibliographystyle{IEEEtran}
\bibliography{WSproximity}

\end{document}